\documentclass[aps,pre,twocolumn,amsmath,amssymb,notitlepage,noeprint,nofootinbib,superscriptaddress]{revtex4-1}
\usepackage{graphicx,color,upgreek}
\usepackage{mathtools}
\usepackage{amsmath}
\usepackage{layouts}
\usepackage{float}
\definecolor{darkblue}{rgb}{0,0,0.6}
\usepackage[colorlinks,linkcolor=darkblue,citecolor=darkblue,urlcolor=darkblue]{hyperref}

\usepackage{makecell}
\usepackage[abs]{overpic}
\usepackage{helvet}
\usepackage{graphicx,color,natbib}
\usepackage{placeins}
\usepackage[dvipsnames]{xcolor}
\usepackage{array, booktabs, makecell}
\usepackage{siunitx}
\usepackage{longtable}
\usepackage{tabularx}
\usepackage{booktabs}
\let\oldAA\AA
\renewcommand{\AA}{\text{\normalfont\oldAA}}
\usepackage{textcomp}
\usepackage{color}
\usepackage{amssymb,amsmath} 
\usepackage{siunitx}

\makeatother
\begin{document}

\preprint{AIP/123-QED}

\title{Anti-correlation between excitations and locally-favored structures in glass-forming systems}

\author{Danqi Lang}
 \email{danqi.lang@ens.psl.eu}
\affiliation{Gulliver UMR CNRS 7083, ESPCI Paris, Universit\'{e} PSL, 75005 Paris, France}
\affiliation{D\'epartement de Physique de l'Ecole Normale Sup\'{e}rieure, ENS, Universit\'{e} PSL, 75005 Paris, France}

\author{Camille Scalliet}
\affiliation{Laboratoire de Physique de l’Ecole normale sup\'erieure, ENS, Universit\'e PSL, CNRS, Sorbonne Universit\'e, Universit\'e Paris Cit\'e, 75005 Paris, France}

\author{C. Patrick Royall}
\affiliation{Gulliver UMR CNRS 7083, ESPCI Paris, Universit\'{e} PSL, 75005 Paris, France}

\date{\today}

\begin{abstract}
Dynamics that are microscopic in space and time, in which particles commit to a position, so-called \emph{excitations}, are considered the elementary unit of relaxation in the Dynamic Facilitation (DF) theory of the glass transition. Meanwhile, geometric motifs known as locally favored structures (LFS) are associated with vitrification in many glassformers. Recent work indicates that the probability of particles found both in locally favored structures (LFS) and excitations decreases significantly upon supercooling suggesting that there is an anti-correlation between them [Ortlieb \emph{et al,} \emph{Nature Commun.} \textbf{14},  2621 (2023)]. However, the spatial relationship between excitations and LFS remains unclear. By employing state-of-the-art GPU computer simulations and colloid experiments, we analyze this relationship between LFS and excitations in model glassformers. We demonstrate that there is a spatial separation between the two in deeply supercooled liquids. This may be due to the fact that LFS are well-packed, thus they are relatively stable. 
\end{abstract}

\maketitle

\section{Introduction}
\label{sectionIntroduction}

The process of vitrification, whereby a liquid solidifies without crystallizing, remains a major challenge in condensed matter. There are a variety of theories postulated, which provide equally good descriptions of the observed dynamic slowdown of some fourteen orders of magnitude in relaxation time with respect to the normal liquid~\cite{berthier2011}. Some are even understood to be incompatible, for example those which relate the dynamic slowdown to a thermodynamic transition to a putative amorphous state with sub--extensive configurational entropy known as the ideal glass~\cite{adam1965,lubchenko2007} while others suppose that the glass transition is a dynamical phenomenon~\cite{chandler2010,speck2019}. Still others relate the glass transition to the emergence of geometric motifs associated with local order.
These so-called locally favored structures (LFS) may be  amorphous~\cite{tarjus2005,royall2015physrep} or crystalline~\cite{leocmach2012}.

Particle--resolved studies, using computer simulation or optical imaging of colloids~\cite{hunter2012,ivlev,royall2023,gokhale2016advphys} promise the level of data to discriminate between the predictions made by these competing theoretical descriptions. Yet until recently, these are limited to the weakly supercooled regime of around four orders of magnitude increase in relaxation time. Now this regime is well--described by the Mode--Coupling theory~\cite{charbonneau2005}. Particle--resolved data obtained at state points which are more deeply supercooled than the crossover of Mode-Coupling Theory (MCT) is needed to make progress. %discriminate between these theories. 
Recently, such data has become available, using SWAP Monte Carlo~\cite{ninarello2017,scalliet2022},  and GPU processors~\cite{bailey2017,ortlieb2023}. Colloid experiments too have passed the MCT crossover, using smaller particles~\cite{brambilla2009,hallett2018,ortlieb2023}. However, rather than providing clear support for one particular theoretical approach, in fact these new data seem to support \emph{multiple} theories. One of these is dynamic facilitation theory~\cite{chandler2010,speck2019,hasyim2024} which predicts that relaxation occurs by \emph{excitations} which are microscopic in space and time. These are indeed found at the population, size and duration predicted in simulations~\cite{keys2011,ortlieb2023,hasyim2021,hasyim2024} and experiments~\cite{gokhale2014,gokhale2016advphys}. 
On the other hand the  \emph{cooperatively re-arranging regions} (CRRs) predicted by the thermodynamics-based approaches of the Random First--Order Transition theory (RFOT)~\cite{lubchenko2007} and Adam-Gibbs theory~\cite{adam1965} are predicted to grow in size and massively in timescale upon supercooling. These too are  found, consistent with theory and in particular the prediction of their compaction at deep supercooling is  upheld~\cite{nagamanasa2015,ortlieb2023}. In addition, the predicted drop in configurational entropy has been robustly found in a number of studies~\cite{turci2017prx,berthier2017,hallett2018,berthier2019}. It has even been suggested that these two approaches may be reconciled~\cite{royall2020,ortlieb2023}.

One piece of the jigsaw which has received rather little attention is the relationship between locally favoured structures 
and dynamic facilitation theory~\cite{chandler2010,speck2019}. LFS play an important role in frustration based theories~\cite{tarjus2005}. A key piece of evidence that there is at least some structural component to dynamic facilitation comes from the observation that the \emph{time--averaged} population of LFS can drive a \emph{dynamical phase transition} which is a central ingredient of the theory~\cite{chandler2010,speck2012,turci2017prx}. Very recently, a strong anti--correlation between LFS and particles in excitations has been found~\cite{ortlieb2023}. This contrasts with previous work which could only access weaker supercooling and did not significantly pass the mode-coupling crossover~\cite{malins2013fara,royall2015physrep}. Amongst the earlier work, some studies claimed a significant relationship between local order and dynamic heterogeneity~\cite{kawasaki2007,leocmach2012,tamborini2015}, while other work found rather little~\cite{charbonneau2012}. While it is likely that the means of determining the structure may be important~\cite{richard2020,tong2018,Cubuk2015,malins2013tcc}, in any case, structure-dynamics relationships in the weakly supercooled regime in 3$d$ have been found to be model--dependent~\cite{hocky2014}.

The fact that it is now possible to study glass-formers at the single-particle level in real space at deeper supercooling than the Mode-Coupling crossover opens the way to probe the role of LFS in this newly accessible dynamical regime. The anti-correlation between LFS and excitations~\cite{ortlieb2023} motivates us to further investigate the relationship between LFS and excitations of dynamic facilitation, particularly at deeper supercooling. This investigation forms the topic of this work. Moreover, if one imagines that the LFS are somehow stable relative to other parts of the system, then upon supercooling they might be expected to ``expel'' excitations. Here, we investigate this hypothesis using GPU computer simulations of a model glassformer and colloid experiments.

This paper is organised as follows. In Sec. \ref{sectionMethods} we describe our computational and experimental models and methods. We present our results in Sec.~\ref{sectionResults}. Our findings are discussed in Sec.~\ref{sectionDiscussion} and we conclude in Sec.~\ref{sectionConclusion}. Supporting information is provided in Sec.~\ref{sectionSupporting}.

\section{Methods}
\label{sectionMethods}

\begin{figure*}
\includegraphics[width=\textwidth]{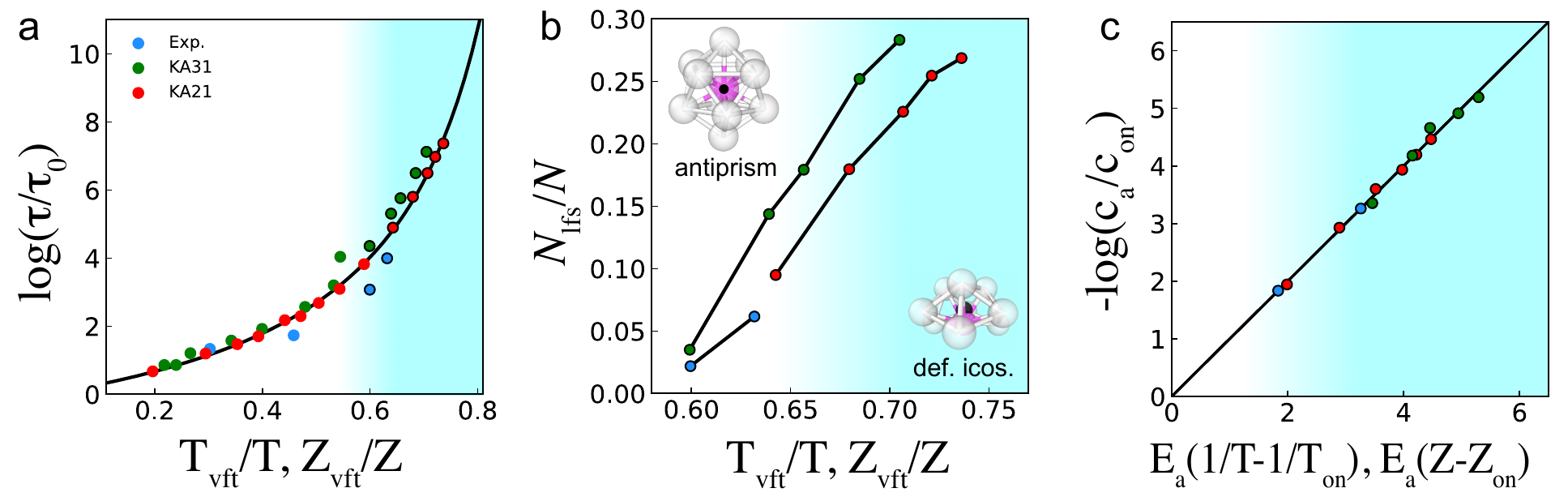} 
\caption{Characteristics of the systems considered.
(a) Relaxation time $\tau_{\alpha}$ with inverse temperature $T$ for Kob-Andersen binary (KA) mixtures (data from~\citet{ortlieb2023}) and compressibility $Z$ for the hard-sphere colloidal system (data from~\citet{royall2018jcp}), scaled with Vogel-Fulcher-Tamman (VFT) parameters given in Table~\ref{tabVFTfit}. The black line is the VFT fit for KA 2:1. We perform our analysis on the data points circled in black, below the mode-coupling crossover (blue shading).
(b) Population of long-lived LFS: the bicapped square antiprism for KA mixtures (top left), and the defective icosahedron for the experimental colloidal system (bottom right). The pink particle is considered the central particle, while the black point is the geometric centre of the LFS.
(c) Scaling of the excitation concentration $c_a$, with the fitting parameters given in Table~\ref{tab:tabEXscaling}.}
\label{figAngell}
\end{figure*}

\subsection{Details of computer simulations}
\label{sectionSimulations}

We study the Kob-Andersen (KA) binary Lennard-Jones mixture at 2:1 and 3:1 compositions with particle density $\rho = 1.4$. The system size is $N$=10002 for the 2:1 composition and $N$=10000 for 3:1. The interactions in the KA mixture are defined by the Lennard-Jones potential
    
\begin{equation}
{u}_{ij}(r)={\epsilon }_{\alpha \beta }\left[{\left(\frac{{\sigma }_{\alpha \beta }}{r}\right)}^{12}-{\left(\frac{{\sigma }_{\alpha \beta }}{r}\right)}^{6}\right]
\end{equation}    
with parameters $\sigma_{AA} = 1$, $\sigma_{AB}=0.80$, $\sigma_{BB}=0.88$, and $\epsilon_{AA} = 1$, $\epsilon_{AB}=1.50$, $\epsilon_{BB}=0.50$  ($\alpha, \beta=A, B)$. We employ $m_{A}=m_{B}=1$. Times are expressed in units of $\sigma_{AA} \sqrt{m_A/\epsilon_{AA}}$.

Computer simulations were carried out using Roskilde University Molecular Dynamics (RUMD). %^ implementation. 
This is a molecular dynamics code which takes advantage of multiple GPU cores to achieve high performance~\cite{bailey2017}. RUMD makes it possible to probe the system at very low temperatures in equilibrium conditions (e.g. $T = 0.48$, where the structural relaxation time $\tau_{\alpha}>10^6$). More details on the RUMD simulations are provided in ~\cite{ortlieb2023}.

We simulate the systems in equilibrium conditions at temperature $T=$ 0.48, 0.49, 0.50, 0.52, 0.55 for the 2:1 composition, and $T =$ 0.68, 0.7, 0.73, 0.75, 0.8 for the 3:1. Once thermalization is reached at these temperatures, we use the LAMMPS package~\cite{plimpton1995,thompson2022} to produce short trajectories of 1000 LJ time units. We sample configurations at a time interval of 1 LJ time unit along the trajectory in order to identify excitations and LFS. For each temperature, we produce 10 independent trajectories, over which we average results.

We employ the conjugate gradient method to find the \emph{inherent structure} (IS) of each sampled configuration. Our analysis is carried out using the IS trajectories, as we found this choice provides a stronger signal than the thermalized configurations.
%instantaneous/thermal ones.

\subsection{Experimental details}
\label{sectionExperimental}

We carry out confocal microscopy experiments with colloidal particles which we track at the single particle level in space and time. These particles closely approximate the hard sphere model~\cite{royall2023,royall2018jcp}.  We use fluorescently labeled density and refractive index matched colloids of sterically stabilized polymethyl methacrylate. The diameter of the colloids is $\sigma^\mathrm{exp}$= 3.23 $\mu$m and the polydispersity is 6\% which is sufficient to suppress crystallization. The Brownian time to diffuse a radius is $\tau_B=6.06$ s. The particles were labelled with the fluorescent dye 3,3’-dioctadecyloxacarbocyanine perchlo- rate (DiOC$_{18}$). Further details are available in~\citet{royall2018jcp}.  In this system, the locally favoured structure is the defective icosahedron~\cite{royall2015,royall2018jcp}. This structure is pictured in Fig.~\ref{figAngell}(b).

The glass transition of hard spheres is obtained via compression, or increasing the volume fraction $\phi$. A convenient quantity to express the state point is the reduced pressure $Z=\beta P/\rho$~\cite{berthier2009}, with $\beta=1/k_BT$ the inverse thermal energy, $P$ the pressure and $\rho$ the number density. Here $Z$ is determined from the Carnahan--Starling relation

\begin{equation}
Z=\frac{1+\phi+\phi^2-\phi}{(1-\phi)^3}~~.
\end{equation}

\noindent Although the colloidal system approaches its glass transition via compression rather than cooling, to facilitate comparison with the simulations and, more generally, molecular systems, we still refer to the colloidal system as being supercooled~\cite{royall2023}.

We consider two state points, $\phi$=0.587 and $\phi$=0.593, corresponding to $Z$= 24.54 and $Z$=25.75. The trajectories last 421$\tau_B$ and 4270$\tau_B$, respectively.

\subsection{Detecting excitations}
\label{sectionExcitation}

We first describe the procedure to detect excitations in simulations and experiments, based on the algorithm proposed by Ortlieb \emph{et al.} ~\cite{ortlieb2023}. %al. takes a point
We analyse molecular dynamics data applying this algorithm to the inherent state trajectory, instead of the thermal one. To analyze experimental trajectories, we make a few small changes as described in the procedure below. The algorithm to detect excitations proceeds as follows:

\begin{enumerate}

\item \emph{Dividing the trajectory}. We divide each 1000 LJ time unit trajectory into 5 sub-trajectories of length $t_a$. We use $t_a = 200$ LJ time units for simulations, and $t_a=200 \tau_B$ for experiments.
        
\item \emph{Testing whether the particle has committed to a new position.} For each particle, we compare its average position in the first and last sub-trajectories. If the difference between these positions is smaller than a predefined threshold $a$, the particle is rejected, \emph{i.e.} not considered an excitation. We choose $a = 0.5\sigma$. 

\item \emph{Determining the time at which the excitation takes place.} This time is denoted as $t_0$, the excitation duration $\Delta t$ and its displacement $\Delta x$. They are found using a hyperbolic tangent fit to the particle position as a function of time. We set a sliding window on which the fit is applied, initially considering every frame as the center of the time window (corresponding to a potential jump in particle position, equivalently the excitation location). The excitation time $t_0$ corresponds to the center of the time window for which the fit is locally optimal. We find that there is a small probability ($<5\%$) that one particle exhibits multiple excitations. Thus, if two locally optimal fits yield excitation times $t_0$ separated by more than $t_a$, we keep both as excitations.
         
\item \emph{Excluding slow and steady movement.} We do not consider as excitations particles that yield a duration $\Delta t > \frac{3}{4} t_a$ and displacement $ \Delta x < a$. 

\item \emph{Excluding the return of particles to the original position.} For the simulations, we continue the MD trajectory for another 1000 LJ time units and check the final displacement of those particles. More precisely, we compare their average position in the first and last $t_a$ of the total 2000 LJ time units trajectory. We exclude particles with a displacement smaller than $a$. For the experiments, we consider the remainder of the trajectory.
        
\end{enumerate}

Particles which satisfy these criteria are considered to be in an excitation. 

In addition to identifying individual excitations along trajectories, we also compute the concentration of excitations $c_a$ defined as the number of excitation found per trajectory (containing 1000 configurations).

\subsection{Identifying long-lived LFS}
\label{sectionIdentifyingLFS}

We now turn to the identification of LFS. We use the Topological Cluster Classification (TCC) algorithm to probe LFS in the Kob-Andersen mixtures and hard sphere colloids~\cite{malins2013tcc}. The LFS has been previously identified for the Kob-Andersen model as the bicapped square antiprism and defective icosahedron for hard spheres ~\cite{coslovich2007, malins2013fara, crowther2015, dunleavy2015, royall2015}. These LFS are comprised of 11 and 10 particles respectively and are illustrated in Fig.~\ref{figAngell}(b). The initial step of the TCC algorithm involves identifying bonds between neighboring particles. These bonds are detected using a modified Voronoi method, where a maximum bond length cutoff of $r_c = 2.0$ is applied for all types of interactions (AA, AB, and BB). Additionally, for KA the parameter $f_c$ is set to unity to control the identification of four-membered rings as opposed to three-membered rings, while $f_c=0.82$ for hard spheres and all particles are treated equally (we neglect polydispersity, following previous work~\cite{malins2013fara, dunleavy2015, royall2015}).

After identifying the LFS, we are interested in selecting only those that are long-lived. In simulations, we define as long-lived the LFS that survive over 80\% of the duration of the trajectory, \emph{i.e.} longer than 800 LJ time units, not necessarily consecutively. We choose the 80\% criterion on the basis of the probability distribution of LFS lifetime, which serves as a turnover point, see Fig. \ref{SFigLFSlifetime}(a,b). For experiments, we consider clusters that survive over 40\% time of the whole trajectory to be long-lived LFS, according to the LFS lifetime distribution Fig. \ref{SFigLFSlifetime}(c).

In the following, our analysis is based on long-lived LFS.

\begin{figure}[t]
\includegraphics[width=60mm]{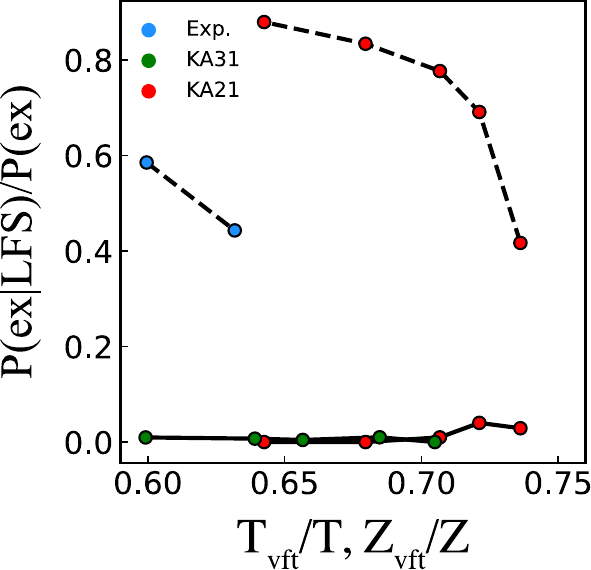}
\caption{Anti-correlation between LFS and excitations. Thermalized/instantaneous data from~\citet{ortlieb2023} (black dashed line with red data points). Experimental data with long-lived LFS is shown with the blue data points. Solid lines correspond to the anti-correlation between long-lived LFS and excitations in the inherent states. 
}
\label{figAntiCorrelation}
\end{figure}

\subsection{Calculation of the separation between LFS and excitations}
\label{sectionOverlap}

If a particle is found to belong to a long-lived LFS, we label it as ``LFS'' throughout the whole trajectory. For excitations, we label the particle as ``EX'' during $[t_0-t_a,t_0+t_a]$, where $t_0$ is the time at which the excitation occurs. We choose these different time criteria between the LFS and excitations because in the case of an excitation occurring \emph{inside} an LFS, its movement would naturally distort the LFS and likely lead to the LFS no longer being detected. If the LFS is no longer detected, then its separation from the excitation would not be counted, which would have severe consequences for our analysis. We have found that the method we implemented avoids this problem.

From this labelling, we then calculate the conditional probability that a particle is in an excitation given that it belongs to an LFS $P(EX \vert LFS)$, and the probability of finding excitations $P(EX)$, both computed in each configuration along the trajectory. The ratio $P(EX \vert LFS) / P(EX)$ thus indicates the probability of excitations overlapping with LFS. For each temperature, we compute the average of that ratio over the 10 independent trajectories, each comprising 1000 configurations.

\begin{figure*}
\includegraphics[width=\textwidth]{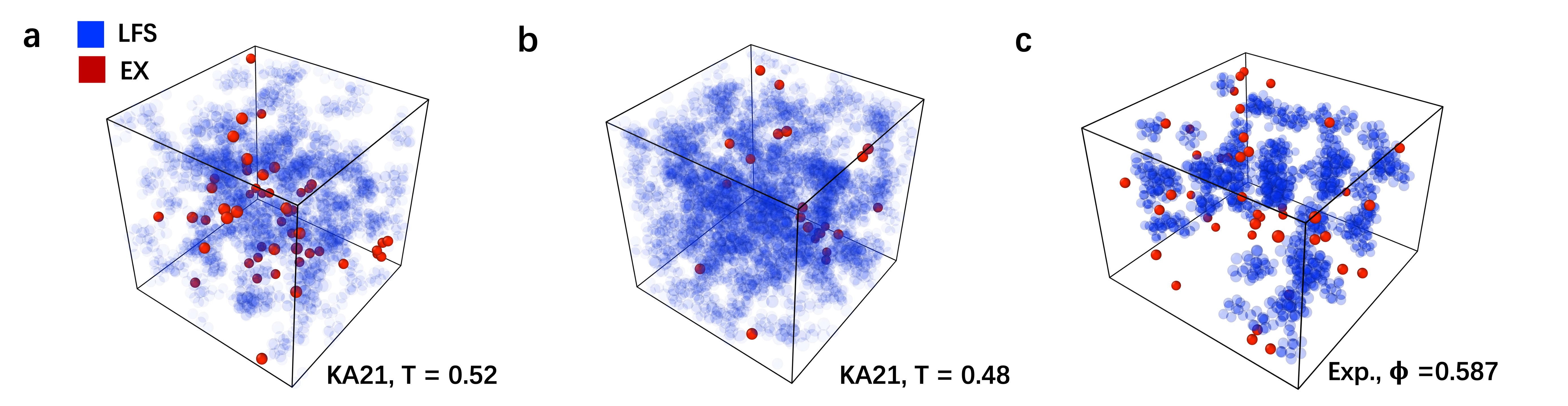}
\caption{Representative snapshots of high (a) and low (b) temperature configurations in KA 2:1. Particles in LFS are rendered in blue and excitations are rendered in red. (c) A representative snapshot of the experimental colloidal system.
}
\label{figOverlap}
\end{figure*}

With this labelling method, we compute the distance between excitations and their closest LFS ``center''. We consider as the LFS center the particle closest to the geometric center of the cluster. In the insets of Fig.~\ref{figAngell}(b), we highlight the particle considered as ``center'' LFS in pink, as well as the location of the geometric center with a black dot. For the antiprism, the central particle coincides with the geometric center. For the defective icosahedron, we take the particle rendered in pink, which is slightly different (0.45$\sigma$ away) from the geometric center. Then we compute the average distance of all excitations in all configurations and trajectories to the center of their closest LFS.

\section{Results}
\label{sectionResults}

We begin our results section by discussing the glass-forming behavior of the systems considered. We probe the increase in relaxation via an Angell plot, and evaluate the increase in LFS upon supercooling. We also determine the drop in fraction of particles in excitations with supercooling. We then move on to consider the spatial relationship of LFS and excitations, starting with some snapshots, before presenting our protocol to determine the separation between excitations and LFS. We then investigate the local packing of particles in excitations and LFS.

\subsection{The proportion of LFS and excitations at deep supercooling}
\label{sectionproportion}

Figure~\ref{figAngell}(a) shows a so-called Angell plot of relaxation time $\tau_{\alpha}$ with respect to inverse temperature $1/T$ for the Kob-Andersen binary (KA) mixture and reduced pressure $Z$ for the experimental hard-sphere colloidal system. Here we fit $\tau_{\alpha}$ with the Vogel–Fulcher–Tamman (VFT) equation,

\begin{equation}
\label{eqVFT}
    {\tau }_{\alpha }^{\mathrm{\\sim}}={\tau }_{0}\exp \left(\frac{D{T}_{{{{{{{{\rm{vft}}}}}}}}}}{T-{T}_{{{{{{{{\rm{vft}}}}}}}}}}\right)\quad, 
    {\tau }_{\alpha }^{\exp }={\tau }_{0}\exp \left(\frac{D{Z}}{{Z}_{{{{{{{{\rm{vft}}}}}}}}}-Z}\right)
\end{equation}

\noindent
for simulations and experiments, respectively. Here $D$ is a system-dependent constant. We focus on the regime at deeper supercooling than the mode-coupling crossover as indicated by the blue shading. The mode-coupling crossover is taken from~\citet{ortlieb2023} to be $T_\mathrm{mct}=0.55\pm0.09$ and $0.7\pm0.1$ for the KA 2:1 and 3:1 systems respectively.  For the experiments, we take a volume fraction $\phi_\mathrm{mct}=0.58$~\cite{vanmegen1998} which corresponds to a reduced pressure $Z_\mathrm{mct}=23.23$.

We now turn to the evolution of the population of LFS, defined as the number of long-lived LFS $N_\textrm{lfs}$ divided by the number of particles $N$. In Fig.~\ref{figAngell}(b) we see that the population of LFS grows with supercooling in all systems. At the lowest temperature, one particle in three belongs to a long-lived LFS.

We define the concentration of excitations $c_a$ as the fraction of particles identified as excitations during the whole trajectory. Dynamic facilitation theory predicts it to decrease exponentially with the inverse temperature~\cite{keys2011} which was verified at deeper supercooling in Ref.~\cite{ortlieb2023}. We show in Fig.~\ref{figAngell}(c) that the data obtained on the IS trajectories is consistent with such scaling.

\begin{figure}
\includegraphics[width=80 mm]
{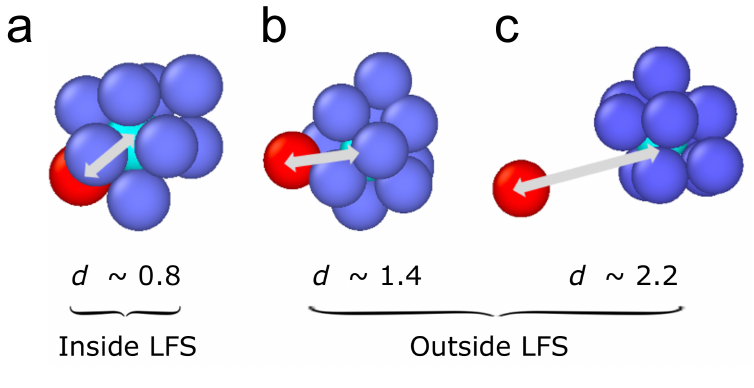}
\caption{Renderings of some typical examples of excitations inside LFS (a) and outside LFS (b,c), with $d$ the distance from an excitation particle (red) and its closest LFS center (cyan). }
\label{figDistanceKASnap}
\end{figure}

\begin{figure*}
\includegraphics[width=\textwidth]{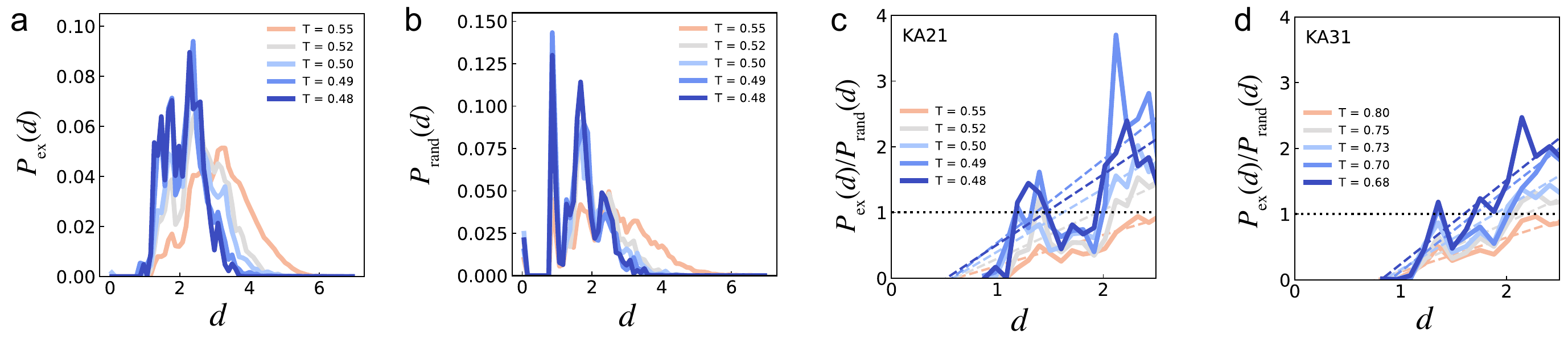}
\caption{The spatial relationship between excitations and LFS for the computer simulations.
Shown is the probability distribution of the distance $d$ from the LFS center as expressed in Fig.~\ref{figDistanceKASnap}.
We consider the following cases.
(a) Excitation to closest LFS center.
(b) A randomly chosen particle to the closest LFS center.
(c) The ratio of (a) and (b). That is to say, normalizing the excitation--LFS center distance by a randomly chosen particle--LFS distance, which accounts the increase in LFS population with supercooling. Dashed lines are linear fits.
(d) Normalized excitation--LFS distance for the KA 3:1 mixture.
}
\label{figDistanceKAGraf}
\end{figure*}

\subsection{The anti-correlation between LFS and excitations}
\label{sectionoverlap}

We further calculate the probability of overlap between LFS and excitations. It turns out that this probability is generally very small (Fig. \ref{figAntiCorrelation}). Representative snapshots shown in Fig. \ref{figOverlap} also confirm that LFS and excitations generally do not overlap, indicating that there may be a spatial separation in between.

An anti-correlation between LFS and excitations was found by~\cite{ortlieb2023}. That work considered thermalized configurations. Here we show that the anti-correlation is robust to inherent states and we also consider experimental data. In fact, while the thermalized configurations with instantaneous LFS show a weaker anti-correlation at higher temperatures, the long-lived LFS considered here used here have a strong anti-correlation at all temperatures as we see in Fig.~\ref{figAntiCorrelation}. We return to this point below.

\subsection{The spatial separation between LFS and excitations}
\label{sectionSpatial}

\begin{figure*}
\includegraphics[width=125mm]{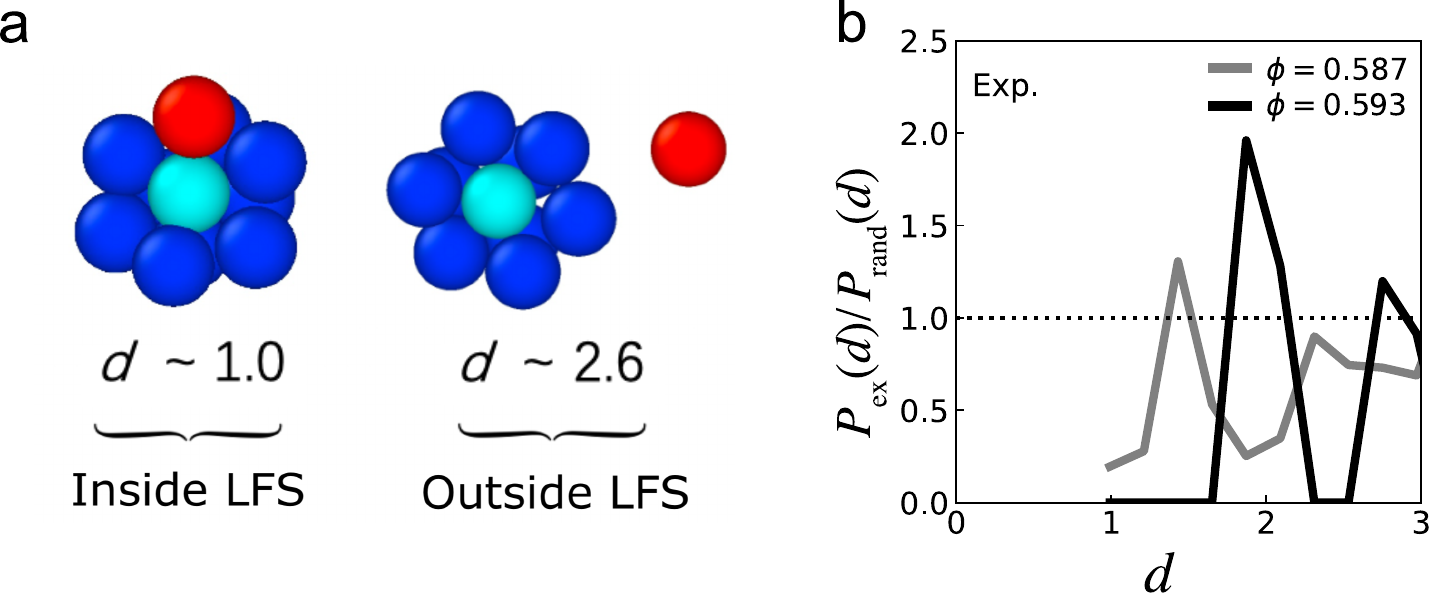}
\caption{The spatial relationship between excitations and LFS for the experiments.
(a) Some typical scenarios of particles inside LFS and outside LFS, with $d$ the distance from one particle and its closest LFS center. 
(b) Scaled excitation--LFS center distance. This corresponds to Figs.~\ref{figDistanceKAGraf}(c,d). 
}
\label{figDistanceExp}
\end{figure*}

To further probe the anti-correlation between LFS and excitations and their spatial separation, we calculate the distance between LFS and excitations. We compute the distance, noted $d$, of an excitation particle to its closest LFS center. Figures~\ref{figDistanceKASnap} and~\ref{figDistanceExp}(a) render some typical scenarios where the excitation particle is inside or outside an LFS, for simulations and experiments respectively.

We compute the probability distribution $P_\mathrm{ex}(d)$, shown in Fig. ~\ref{figDistanceKAGraf}(a). However, since the population of LFS increases with supercooling [Fig.~\ref{figAngell}(b)], one expects that the distance between a particle chosen at random and its nearest LFS particle to \emph{decrease} with supercooling. This is the opposite of our hypothesis of an \emph{increasing} separation between LFS and excitations. This effect seems to lead to a \emph{decrease} in typical separation between LFS and excitation particles as shown in Fig.~\ref{figDistanceKAGraf}(a).

To compensate for this effect of the decrease in excitation--LFS separation, which we expect to be driven by the increase in LFS particles with supercooling, we proceed as follows. We pick some random particles and also compute the probability distribution $P_\mathrm{rand}(d)$ of its distance to the closest LFS center, which we show in Fig.~\ref{figDistanceKAGraf}(b). This plot is related to the pair distribution function, but constrained to one particle being a LFS center. It therefore exhibits peaks at multiples of the particle diameters. However, it is cut-off at large distance because one considers the \emph{closest} LFS center. This cutoff value decreases with temperature, which is consistent with the increase in LFS density reported in Fig.~\ref{figAngell}(b). 

By normalizing the excitation-LFS separation distribution by that of the random particles-LFS, we can compensate for the effect of the increasing LFS population at deep supercooling. Thus by calculating $P_\mathrm{ex}(d) / P_\mathrm{rand}(d)$, we obtain Fig.~\ref{figDistanceKAGraf}(c), which reveals the distribution of distances between LFS centers and excitations
compared with a random particle. The horizontal dotted line $P_\mathrm{ex}(d) / P_\mathrm{rand}(d) = 1 $ in Fig.~\ref{figDistanceKAGraf}(c) separates spatial regions where the probability of finding an excitation is smaller or greater than finding a random particle. We see that the probability to have an excitation inside the LFS, corresponding to $d<1$ or Fig.~\ref{figDistanceKASnap}(a), is zero. Instead, we observe that excitations are increasingly `expelled' from the center of LFS as temperature decreases. We indeed identify two peaks above the horizontal line, located around $d \sim 1.4$ and $d \sim 2.2$, with the second peak higher and increasing with supercooling, indicating a larger probability to find excitations further away from LFS. The location of these peaks correspond to excitation-LFS distances shown in Fig.~\ref{figDistanceKASnap}(b) and (c), respectively.

In order to highlight this increasing spatial separation between excitations and LFS, we fit the ratio of probabilities by a linear function, shown with dashed lines. We find that there is a trend that excitations are more likely to occur outside LFS, and even so at low temperature. This spatial separation between LFS and excitations is also observed in the KA 3:1 mixture [Fig.~\ref{figDistanceKAGraf}(d)] and experimental colloidal system (Fig.~\ref{figDistanceExp}). Thus, for all systems considered here, there is a tendency towards excitations occurring increasingly further away from LFS upon supercooling. This is the main result of our study.

\subsection{Voronoi Cell Volumes}
\label{sectionVoronoi}

\begin{figure}
\includegraphics[width=\columnwidth]{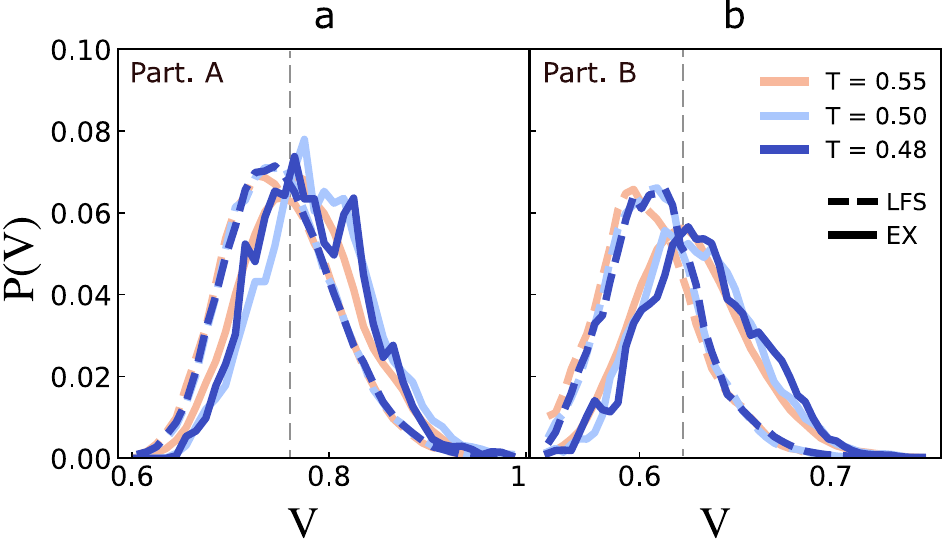}
\caption{Volumes of Voronoi cells for particles in LFS and excitations in the KA 2:1 composition. The probability distribution of (a) A particles, and (b) B particles. The vertical black line is the average over all (A or B) particles. Solid lines are particles in excitations, dashes are particles in LFS. Data for the KA 3:1 mixture is provided in Fig.~\ref{SFigKA31Voronoi}.
}
\label{figVoronoi}
\end{figure}

To investigate why excitations are more likely to occur outside LFS, we compute the volume of the Voronoi cell for each particle. We consider A and B particles for the KA 2:1 composition Fig.~\ref{figVoronoi}, where we plot the distribution of Voronoi volumes for particles in LFS and also those in excitations. While the distributions overlap, there is a clear trend of LFS particles having a small Voronoi volume, \emph{i.e.} being well packed. The opposite trend is seen for excitations whose larger Voronoi volumes indicate poorer packing. In the case of the smaller B particles, the trend is stronger. We show the case for the KA 3:1 composition in the Supplementary information, see Fig.~\ref{SFigKA31Voronoi}. For the experimental data, errors in coordinate tracking hamper this analysis, moreover for a given coordinate, we do not know the diameter of a given particle in this polydisperse system~\cite{ivlev,royall2023}.

\section{Discussion}
\label{sectionDiscussion}

We have investigated the spatial relationship between LFS and excitations in simulations and experiments on model glassformers. This work was motivated by the recent observation of a significant %strong 
anti-correlation between excitations and LFS~\cite{ortlieb2023}. This seems to be stronger than is the case for earlier work which examined the relation between local structure and dynamic heterogeneity at weaker supercooling~\cite{charbonneau2012,malins2013fara}.

Here we found a significant anti-correlation between \emph{long-lived} LFS and excitations. This is stronger than that observed previously~\cite{ortlieb2023}, due to our use of long-lived LFS and using inherent state coordinates. Some comments are in order here. In a series of papers Speck and co-workers explored the dynamical phase transition of dynamic facilitation theory in a number of model glassformers in both simulations and experiment~\cite{speck2012,pinchaipat2017,turci2017prx,turci2018,royall2020,campo2020}. This transition -- obtained using short (a few $\tau_\alpha$) and small (typically around 100-200 particles) trajectories -- exists between a so-called active phase (similar to the normal supercooled liquid) and an inactive phase, whose dynamics are too slow to be measured on the simulation timescale. In the work of Speck and collaborators, this transition is driven by the \emph{time--averaged} population of LFS, (or by dynamics, as obtained previously~\cite{hedges2009,speck2012jcp}). Therefore, the anti-correlation between excitations (a feature of the active phase) and long-lived LFS (which are likely related to the time-averaged LFS of the inactive phase) may be taken as consistent with the dynamical phase transition. We therefore expect that long-lived LFS are some measure of local stability in the supercooled liquid.

The deeper supercooling now possible in both experiment and computer simulation enables us to identify excitations (which are hard to detect for $T>T_\mathrm{mct}$ and $Z<Z_\mathrm{mct}$). Earlier work often struggled to find a significant coupling between local structure and dynamic heterogeneity~\cite{charbonneau2012,malins2013fara}.

It has been found that there is a rather weaker anti-correlation between LFS and cooperatively re-arranging regions (CRRs) which are the elementary units of relaxation in RFOT and Adam-Gibbs thermodynamic-based theories~\cite{ortlieb2023}.
If we imagine that the long-lived LFS are representative of the emerging solid glass, the enhanced anti-correlation of the excitations with respect to the CRRs raises questions about these relaxation mechanisms. It has been suggested that CRRs may be comprised of many excitations~\cite{ortlieb2023}. The lifetimes even of the long-lived LFS are very much smaller than the structural relaxation time $\tau_\alpha$ and indeed the timescale associated with CRRs, which reaches $1.1\times 10^5$ LJ time units for the lowest temperature studied~\cite{ortlieb2023}. Such a timescale of course presents great challenges for the frequency of sampling that we use here. Here, we have focused on the excitations, but the link between excitations and the much longer timescale CRRs, and the LFS remains an intriguing and challenging topic for a future investigation.

\section{Conclusion}
\label{sectionConclusion}

In this work, we have analysed particle-resolved data from both GPU simulations and experiments in deeply supercooled liquids, with a focus on detecting excitations and identifying \emph{long-lived} locally favoured structures. We have shown that with the decrease in temperature or increase in reduced pressure, the proportion of LFS grows while the proportion of excitations decreases. Moreover, a strong and robust anti-correlation between long-lived LFS and excitations has been identified. By considering inherent states in our computer simulations we have obtained a stronger anti-correlation than that previously found~\cite{ortlieb2023}. To further probe this anti-correlation, we have computed the distance between excitations and long-lived LFS and excitations. Notably, excitations are more likely to occur outside long-lived LFS, demonstrating there is a spatial separation between the two. This spatial separation may be attributed to the well-packed nature of long-lived LFS, as revealed by the analysis of the volumes of the Voronoi cells for each particle.

Our work provides a picture of structural relaxation via excitations occurring in regions of the system outside long-lived LFS. We hope this will stimulate further investigations of the relationship between structure and dynamics in the deeply supercooled regime of glass-forming systems which is now accessible.

\section*{acknowledgments}
The authors would like to acknowledge Ludovic Berthier, Thomas Speck and Gilles Tarjus for insightful discussions.
DL gratetully acknowledges \'{E}cole Normale Sup\'{e}rieure 
for financial support.
CPR would like to acknowledge the Agence National de Recherche for the provision of the grant DiViNew. \\

\section*{Data Availability Statement}

Data and codes are available upon reasonable request to the corresponding author.

\section{Supporting Information}
\label{sectionSupporting}
\setcounter{figure}{0}
\setcounter{table}{0}
\renewcommand{\thefigure}{S\arabic{figure}}

\begin{table*}[htbp]
    \centering
    \begin{tabular}{|l|l|l|l|}
    \hline
        ~ & KA 2:1 & KA 3:1 & Experiment \\ \hline
        $T_{VFT}$ or $Z_{VFT}$ & $0.35$ & $0.48$ & $41$ \\ \hline
        $\tau_0$ & $0.09$ & $0.07$ & $0.63$ \\ \hline
        $D$ & $2.68$ & $2.99$ & $2.21$ \\ \hline
    \end{tabular}
    \caption{VFT parameters fitted to the various systems considered.}
    \label{tabVFTfit}
\end{table*}

\begin{table*}[htbp]
    \centering
    \begin{tabular}{|l|l|l|l|}
    \hline
        ~ & KA 2:1 & KA 3:1 & Experiment \\ \hline
        $T_{on}$ or $Z_{on}$ & $0.75$ & $1.2$ & $23$ \\ \hline
        $E_a$ & $5.96$ & $8.30$ & $1.19 $ \\ \hline
    \end{tabular}
    \caption{Scaling of the fraction of particles in excitations.}
    \label{tab:tabEXscaling}
\end{table*}

\begin{figure*}
\includegraphics[width=\textwidth]{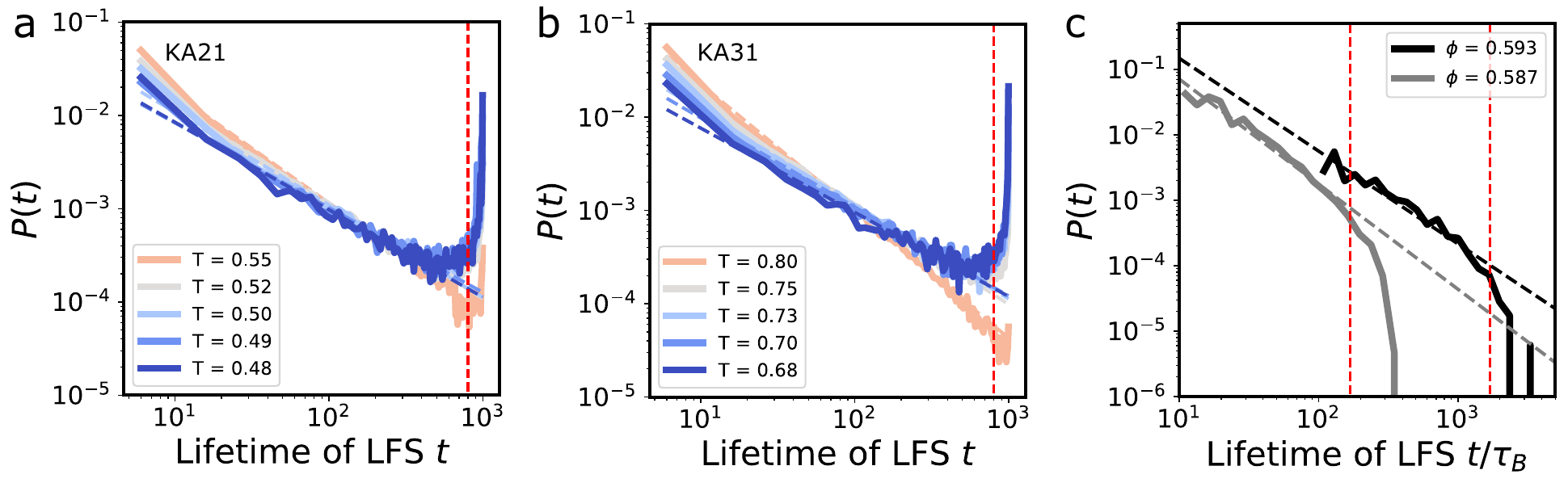}
\caption{The probability distribution of LFS lifetime in simulations of the KA system at (a) 2:1, (b) 3:1 composition, and (c) colloidal experiments. LFS with a lifetime greater than the vertical dashed red lines are defined as long-lived.
}
\label{SFigLFSlifetime}
\end{figure*}

\begin{figure*}
\includegraphics[width=125mm]{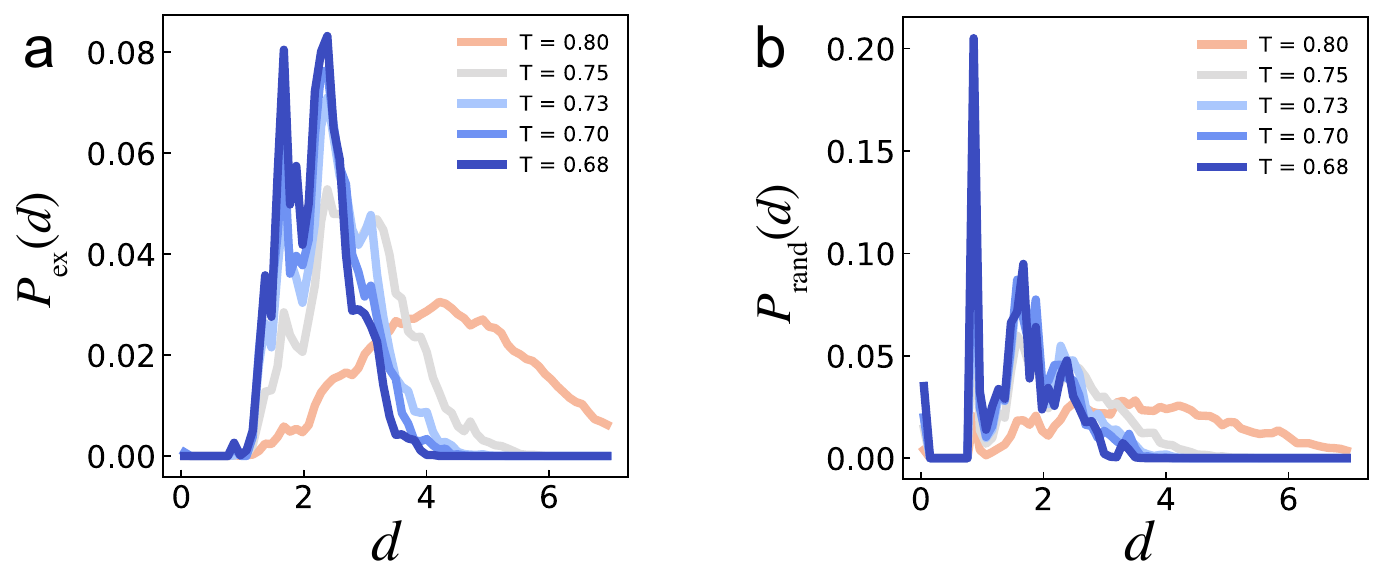}
\caption{The spatial separation in the KA 3:1 system. (a) The probability distribution of the distance $d$ between excitations and their closest LFS centers. (b) The probability distribution of the distance $d$ between random particles and their closest LFS centers. The ratio of the data plotted in (a) and (b) is shown in Fig.~\ref{figDistanceKAGraf}(d) in the main text.
}
\label{SFigDistanceKA31}
\end{figure*}

\begin{figure*}
\includegraphics[width=125mm]{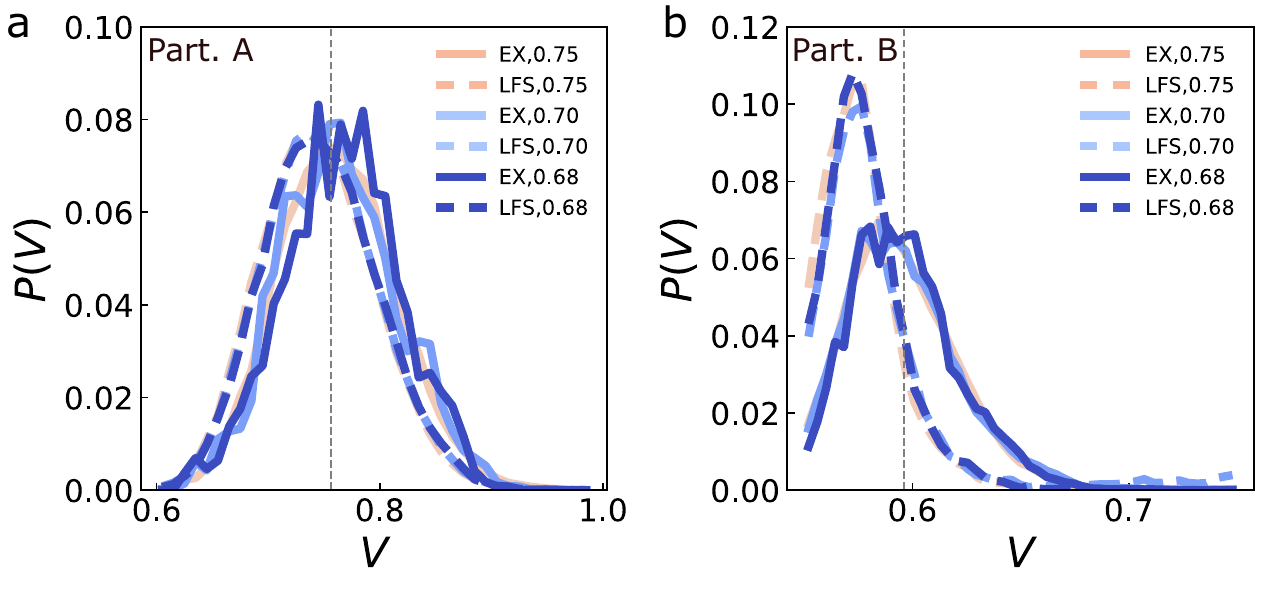}
\caption{Volumes of Voronoi cells for particles in LFS and excitations in the KA 3:1 composition. (a) The probability distribution of A particles. (b) The probability distribution of B particles. The vertical black line is the average of all particles. Solids lines are particles in excitations, dashes are particles in LFS.
}
\label{SFigKA31Voronoi}
\end{figure*}

\newpage

\bibliography{main}

\end{document}